\documentclass[9pt, conference]{IEEEtran}
\usepackage{algorithmicx}
\usepackage{algorithm}
\usepackage{amsmath,amssymb}

\usepackage{algpseudocode}
\usepackage{multirow}
\usepackage{epsfig}
\usepackage{subcaption}
\usepackage{multicol,lipsum,xparse}
\usepackage{epstopdf}

\begin{document}

%\title{\LARGE Advanced Datapath Synthesis using Graph Isomorphism}
\title{Advanced Datapath Synthesis using Graph Isomorphism}

%\numberofauthors{1}
%\author{
%\alignauthor Cunxi Yu, *Mihir Choudhury, *David J. Geiger, *Andrew Sullivan,  Maciej Ciesielski\\
%      \affaddr{University of Massachusetts, Amherst}\\
%       \affaddr{*IBM T.J Watson Research Center}\\
%       \email{ycunxi@umass.edu, choudhury@us.ibm.com}
%}

\author{Cunxi Yu$^1$, Mihir Choudhury$^2$, Andrew Sullivan$^2$,  Maciej Ciesielski$^1$ \\
ECE Department,~University of Massachusetts, Amherst$^1$ \\
IBM T.J Watson Research Center$^2$ \\
    ycunxi@umass.edu, ~choudhury@us.ibm.com}

%\CopyrightYear{2016} 
%\setcopyright{acmcopyright}
%\conferenceinfo{DAC '16,}{June 05-09, 2016, Austin, TX, USA}
%\isbn{978-1-4503-4236-0/16/06}\acmPrice{\$15.00}
%\doi{http://dx.doi.org/10.1145/2897937.2898000}
%
\maketitle
\vspace{-5mm}

{\it\bf Abstract -} 
This paper presents an advanced DAG-based algorithm for datapath synthesis that targets area minimization using logic-level resource sharing. The problem of identifying common specification logic is formulated using unweighted graph isomorphism problem, in contrast to a weighted graph isomorphism using AIGs. In the context of gate-level datapath circuits, our algorithm solves the unweighted graph isomorphism problem in linear time.  The experiments are conducted within an industrial synthesis flow that includes the complete \textit{high-level synthesis}, \textit{logic synthesis} and \textit{placement and route} procedures. Experimental results show a significant runtime improvements compared to the existing datapath synthesis algorithms.


\section{Introduction}

Due to a large demand for computing, the complexity of hardware systems have been significantly increasing, raising the challenges in design, verification and synthesis to a new level. In the last ten years, there has been a push to make changes in optimization algorithms of EDA tools to improve their performance in terms of timing, area and power. Particularly affected are datapath modules in microprocessors and embedded systems which play an important role in computations, which puts new demands on logic synthesis. %However, this is limited to constructability of DFGs and the function of operators. 
Traditional datapath synthesis flow includes extraction of arithmetic operations from RTL code, high-level synthesis (HLS), logic synthesis, and technology mapping \cite{stok1994data}\cite{micheli1994synthesis}. Datapath synthesis techniques have been mainly discussed in the context of traditional high-level synthesis research, such as \textit{resource sharing}, \textit{scheduling} and \textit{binding}, relied on Data Flow Graph (DFG) representation \cite{potkonjak1994optimizing}\cite{srivastava1995optimum}\cite{cong2008simultaneous}. Arithmetic operations such as addition, multiplication, shifting and comparison, and control logic are extracted and modeled as block modules. At the same time, methods such as carry prefix, and recoded partial product based techniques are applied for delay optimization \cite{roy2014towards}. The remaining part of the design flow produces the technology mapped netlist using standard-cell library. 

Even though most of the datapath synthesis effort is spent in the high-level synthesis stage, there are many unexplored opportunities in \textit{bit-level} optimization that could improve results of high-level synthesis. Recently, high-level optimization techniques, such as \textit{resource sharing}, have been applied in logic synthesis to overcome some of the limitation of datapath synthesis for standard-cell designs. Specifically, a Directed Acyclic Graph (DAG) based logic synthesis technique that targets area minimization of datapath designs was proposed in \cite{cunxiyu:dac16}. It is a structural optimization technique implemented using \textit{And-Inv-Graphs} (AIGs) \cite{mishchenko:2006-dag}, which offers bit-level resource sharing. The method includes three steps: 1) identifying sub-circuit candidates by searching a \textit{multiplexer-equivalent} AIG structure; 2) identifying common specification logic using graph isomorphism; and 3) finalizing the optimization by relocating multiplexers across common logic. The most critical part of the technique is step 2, which solves the problems of identifying common logic and performing Boolean matching. In fact, finding isomorphism in AIG is a \textit{weighted} graph isomorphism problem \cite{cunxiyu:dac16}. This is because, to represent an arbitrary Boolean network using AND nodes, the edges are required to represent inversion or a wire, which classifies an AIG as a \textit{weighted} graph. Note that solving graph isomorphism in \textit{weighted} graphs is much more complex than in the \textit{unweighted} graphs \cite{umeyama1988eigendecomposition}.

Although the technique of \cite{cunxiyu:dac16} offers new direction in datapath synthesis and promises area reduction, it has some limitations. First, the complexity of general graph isomorphism problem belongs to \textit{NP}, but is not known if it is \textit{P} or \textit{NP-complete}. Despite the reduction in complexity offered by DAGs, solving a weighted DAG isomorphism could still cause memory and runtime explosion. Furthermore, since that technique is implemented based on AIG, it requires transformations between gate-level network and AIG representations to produce the technology mapped netlist. These transformations could affect the optimization solutions performed by the previous synthesis procedures. %Third, the multiplexer identification method is based on strictly structural searching, that is not general enough for identifying multiplexers, e.g., when vector multiplexers are optimized combinations of selectors and decoders. 

In this work, we develop new algorithms to overcome these limitations. Specifically, we make the following contributions:

\noindent
\textbf{~~1)} We propose a novel algorithm for identifying common specification logic that directly supports arbitrary standard-cell netlist, without using AIG, which maintains the optimizations performed by other synthesis techniques.

\noindent
\textbf{~~2)}  Instead of solving \textit{weighted} graph isomorphism problem, the proposed algorithm formulates the problem as \textit{unweighted} graph isomorphism, which significantly reduces the complexity of solving the problem.

\noindent
\textbf{~~3)}  The runtime complexity comparison between the AIG-based algorithm \cite{cunxiyu:dac16} and the one presented here is provided using illustrative examples (Section 3.1), and demonstrated using large datapath designs (Figure 6).

\noindent
\textbf{~~4)} The proposed algorithm allows \textit{approximate isomorphism classes} to be optimized (Section 3.2).

\noindent
\textbf{~~5)} This approach has been evaluated in two complete IBM synthesis flows, including the complete flow of \textit{high-level synthesis}, \textit{logic synthesis} and \textit{place and route} (P\&R), which allows it to make meaningful comparison with other techniques. The experiments were performed using 14nm technology library.

%\begin{figure}[t] 
%\begin{center}
%\includegraphics[scale=0.32]{motivation.eps}
%\caption{(a) Design flow of datapaths. (b) \textit{Resource sharing} example.}
%\vspace{-3mm}
%\label{fig:motivation}
%\end{center}
%\end{figure}

\section{Background}

\subsection{Boolean Network}
A Boolean network can be represented using directed acyclic graph (DAG) with nodes representing logic gates and directed edges representing wires connecting the gates. If the network is sequential, the memory elements are assumed to be D flip-flops with known initial states. In this work, we only consider combinational logic optimization, which means the flip-flops are considered as primary inputs (PI) and primary outputs (PO) for the sub-circuits.  

In the AIGs \cite{mishchenko:2006-dag}, each node has either 0 or two incoming edges. A node with no incoming edges is a primary input. Primary outputs are represented using special output nodes without output edges. Each internal node represents a Boolean AND function. The combinational logic of an arbitrary Boolean network can be transformed into an AIG \cite{mishchenko2010abc}, while the edges can optionally provide inversions. Hence, AIG is considered as a \textit{weighted} DAG. 

\begin{figure}[t] 
\begin{center}
\includegraphics[scale=0.3]{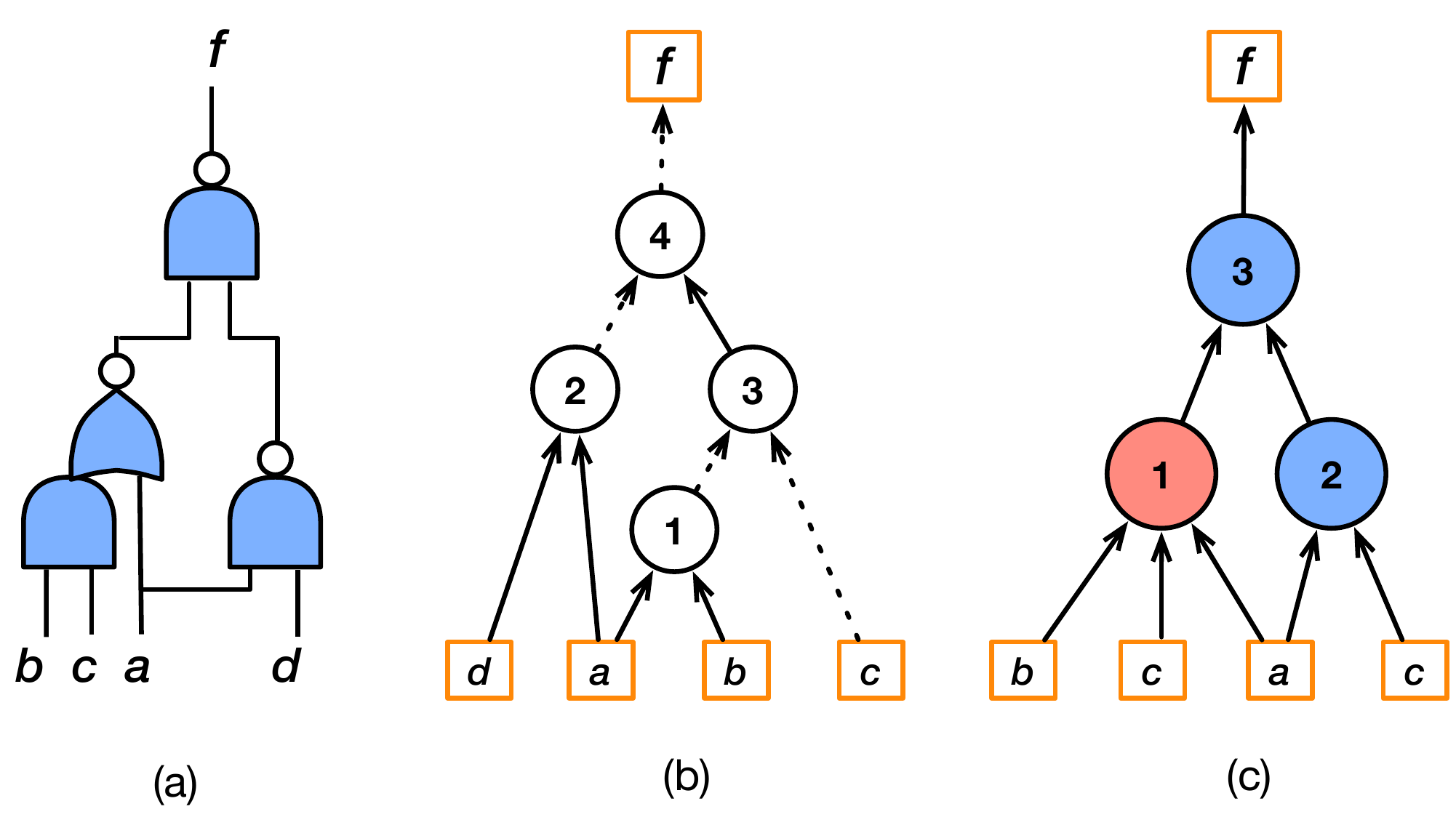}
\caption{(a) Gate-level netlist $f$=$\overline{\overline{bc+a} \cdot \overline{ad}}$; (b) AIG representation, $f$=$\overline{f_2 \overline{f_3}}$, $f_2$=$\overline{f_1}\overline{a}$, $f_{3}$=$ad$, and $f_1$=$bc$; (c) the proposed representation, $node1$ representing AOI21, and $node2$ and $node3$ representing the two NAND2s.}
\vspace{-2mm}
\label{fig:aig_vs_tib}
\end{center}
\end{figure}

Alternatively, the Boolean network can be directly represented using the gate-level netlist. The primary inputs, primary outputs, and flip-flops are constructed based on standard-cell netlist. Each logic gate is a vertex in the DAG. The logic gates with the same corresponding logic function are considered as the same vertex type. This DAG has only one type of edge, i.e. \textit{unweighted} DAG, and provides more uniqueness for checking isomorphism. The comparison between AIG and our representation is shown in Figure \ref{fig:aig_vs_tib}. The actual gate-level netlist, including one AOI21 and two NAND2 gates, is shown in Figure \ref{fig:aig_vs_tib}(a), and its AIG representation is shown in Figure \ref{fig:aig_vs_tib}(b). AIG requires four AIG nodes with four inversion edges and five non-inversion edges to represent this netlist. In contrast, the proposed representation in Figure \ref{fig:aig_vs_tib}(c), has three nodes in two types, and all edges are identical. There are several advantages of the representation shown in Figure \ref{fig:aig_vs_tib}(c) that we adopted in our work: 1) avoid the transformations between different Boolean network to maintain the original structural, which maintains the optimizations done in previous stages; 2) convert the weighted graph isomorphism problem into unweighted graph isomorphism problem to improve the runtime for identifying \textit{common specification logic}. %Two combinational circuits are considered as \textit{common specification logic} if two combinational circuits have common specification. Note that our approach based on this representation supports multiple outputs cells, such as Full adders.

\subsection{Common Specification Logic}

Two combinational circuits are considered as \textit{common specification logic} if they have the same specification \cite{goldberg2004equivalence}. In this work, common specification logic has to be identified in the following context: given the output boundaries of two logic cones, find the input boundaries that result in maximum common logic such that the signals of the input boundaries match. Most techniques for checking if two designs conform to common specification logic are based on \textit{combinational equivalence checking} (CEC). This problem has been addressed by BDDs \cite{bryant1986graph}, SAT\cite{kuehlmann1997equivalence}\cite{goldberg2001using}, AIG\cite{mishchenko2010abc}, etc. However, those methods cannot be applied in this work for the following reasons: 1) the input boundaries of the designs are unknown; and 2) if the input boundaries are detected, the relationship (Boolean matching) of those inputs is unknown. Furthermore, it is well known that the functional methods such as BDDs and SAT, are not scalable for gate-level arithmetic designs, such as multipliers.%This is the reason the proposed approach uses DAG-based structure method. 

%\cite{mishchenko2010abc} is considered as the state-of-art. 

%\subsection{Logic synthesis}
%Logic synthesis transforms the Boolean network to reduce the number of nodes (area), logic levels (delay), switching activity (power). Traditional logic synthesis tools, such as SIS \cite{sentovich1992sis} and ABC \cite{mishchenko2010abc}, target multi-level logic optimization that apply removing redundancy, logic simplification, and logic sharing. Logic synthesis techniques can be classified into functional methods, such as bi-decomposition \cite{yang2002bds}, and structural methods such as logic sweeping \cite{mishchenko2005fraigs}. %Recently, a bottom-up synthesis method was proposed to combine resource sharing in logic synthesis \cite{cunxiyu:dac16}. 
%Technology mapping deals with representing the logic regarding a given set of primitives, such as standard cells or lookup tables. 
%\subsection{Motivation}

\vspace{-4mm}

\subsection{Graph Isomorphism}

In graph theory, an isomorphism of graphs $G$ and $H$ is a bijection between the vertex sets $V(G)$ and $V(F)$, $f$: $V(G)$ $\rightarrow$ $V(F)$, such that any two vertices $u$ and $v$ of $G$ are adjacent in $G$ iff $f(u)$ and $f(v)$ are adjacent in $H$. Besides the mathematical research on graph isomorphism, the algorithmic approach to graph isomorphism has been widely used in computer engineering, e.g. Boolean matching \cite{soeken2015simulation} and program similarity checking \cite{li2016detecting}. In general, graph isomorphism is applicable to undirected, unlabeled, unweighted graphs. Its is known to be an \textit{NP} problem, but neither a \textit{NP-complete} nor a \textit{P} using a deterministic algorithm. However, in the context of Boolean network, this problem could be solved efficiently by heuristic algorithms. In this work, we propose a novel algorithm that reduces the number of reordering operations by employing fanin-fanout information of each node (i.e. standard cell) for checking the existence of an isomorphism between two directed acyclic graphs. %This algorithm mainly utilizes the properties of Boolean network, i.e. directed acyclic and fanin-fanout information of each node (standard cell). %We also study the runtime complexity of the algorithm in \cite{}\

\section{Approach}

The overall methodology of our approach is in three steps. \textit{Vector multiplexer} is a set of 2-to-1 multiplexers with the identical control signals. First, they are collected by first structurally reverse engineering all the 2-to-1 multiplexers from gate-level netlist \cite{cunxiyu:dac16}, and then being classified based on their control signals. Note that the multiplexers are eliminated from the collection if any of their data inputs has a fanout. In case of large multiplexers, such as 64-to-1, they are decomposed into 2-to-1 multiplexers \cite{mitra2000efficient}. Second, a set of sub-circuits is created based on these vector multiplexers. Each sub-circuit is a combinational logic cone whose primary outputs are the outputs of all multiplexers in the vector multiplexer. These two procedures are \textit{pre-processing} step. Third, a multiplexer relocation function is applied to each output of the sub-circuit iteratively. The order of applying multiplexer relocation is sorted by the number of logic gates per multiplexer in the sub-circuit. The original design will be updated if the area of the sub-circuit is improved by relocating the multiplexers, i.e., moving the multiplexers backward without changing the functionality of the design. The resulting updated standard-cell netlist, and will be subjected to the remaining logic synthesis steps and eventually to physical design. %In addition to the main flow, several new heuristics are implemented in this approach.

\subsection{Exact Isomorphism Determination}

Even though the multiplexer relocation is applied to a sub-circuit that includes vector multiplexers at the primary outputs, the actual relocation is done individually for each multiplexer. The goal of multiplexer relocation is to maximize sharing of common specification logic that are the input cones of the multiplexers, by moving the multiplexers backward. The main challenge is to identify the common specification logic in the sub-circuits created by pre-processing step. Specifically, this requires performing common structure identification and Boolean matching. According to the definition of \textit{graph isomorphism}, the algorithm proposed in \cite{cunxiyu:dac16} determines the isomorphism boundary between two graphs using breath-first-search. To obtain the maximum common logic, a look-ahead heuristic is applied in case of there are multiple identical choices of constructing isomorphism. This could potentially cause an exponential runtime and memory explosion problem, especially in the design with many reconvergent fanouts. In this section, we introduce a novel algorithm to improve the runtime and scalability for identifying common specification logic.

\subsubsection{Standard-cell based DAG advantages}

\begin{figure}[!htb] 
\begin{center}
\includegraphics[scale=0.3]{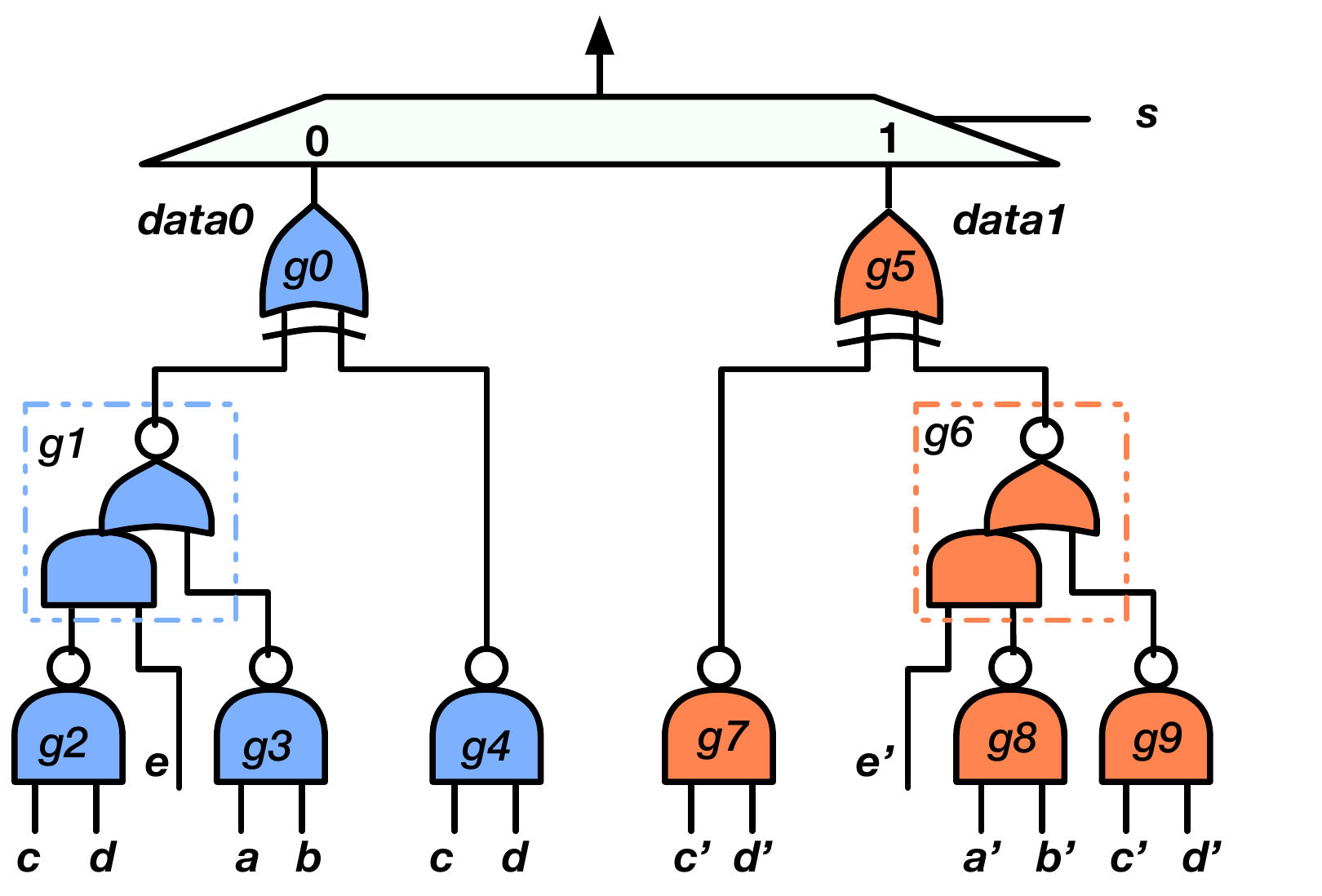}
\vspace{-2mm}
\caption{Determine graph isomorphism using standard-cell based DAG.}
\vspace{-3mm}
\label{fig:tib_exact1}
\end{center}
\end{figure}

\begin{figure}[!htb] 
\begin{center}
\includegraphics[scale=0.3]{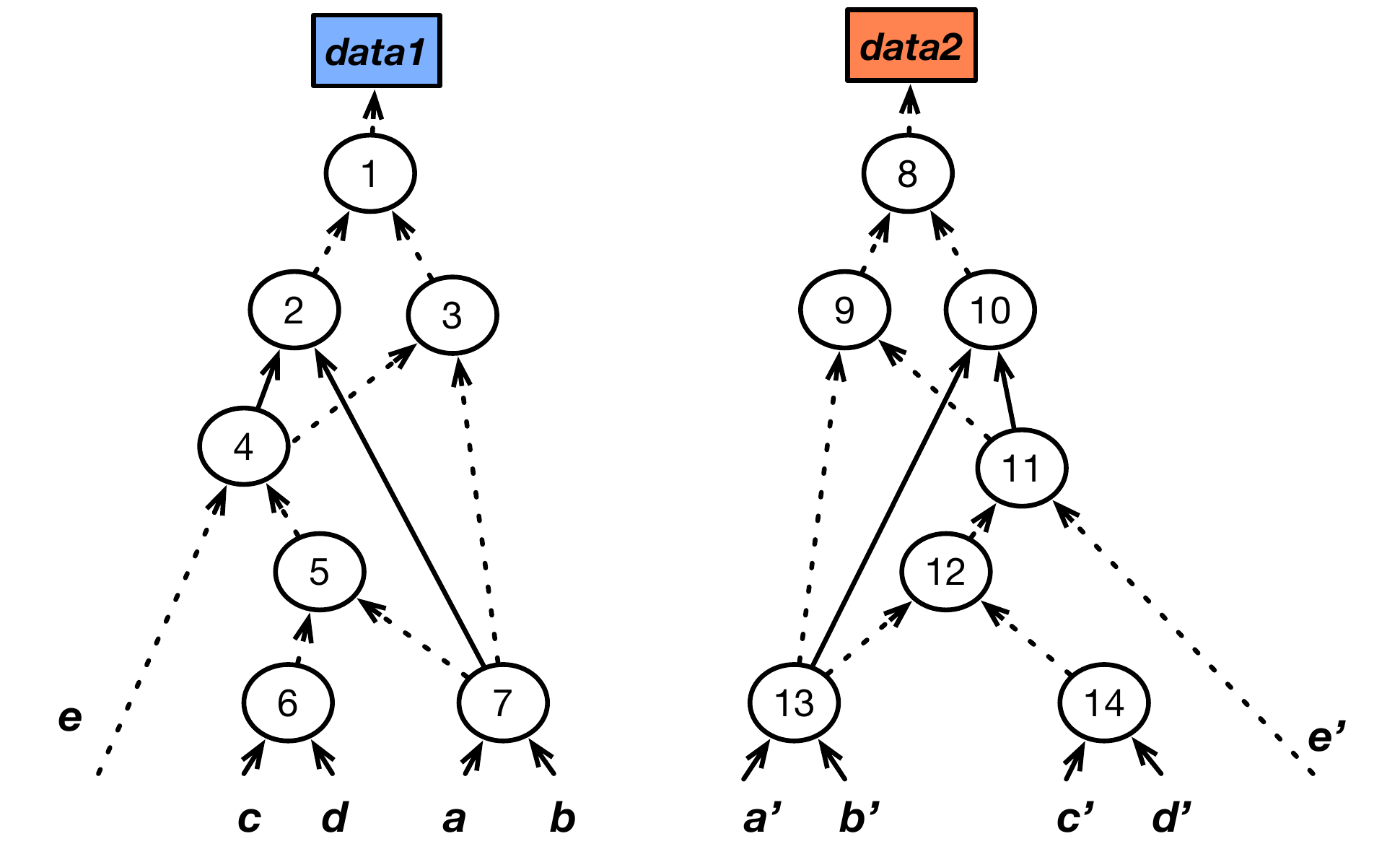}
\vspace{-2mm}
\caption{Determining graph isomorphism using AIG.}
\vspace{-3mm}
\label{fig:tib_exact2}
\end{center}
\end{figure}

Instead of using AIG representation, the standard-cell based representation gives two advantages: 1) some optimization efforts in other stages of the synthesis flow, that may disappear during the transformations between AIG and standard-cell netlist are maintained; 2) standard-cell representation significantly reduces the possible choices for checking the existence of isomorphism. For this advantage, there are three reasons: \textbf{(a)} in each topological level, the total possible pairing choices is reduced; \textbf{(b)} edge type is no longer necessary to be considered, which makes the isomorphism problem to be \textit{unweighted}; and \textbf{(c)} utilizing the number of inputs and outputs of each standard-cell reduces the number of possible choices when checking isomorphism, especially in the representation of logic circuits. We demonstrate these using an example in Figures \ref{fig:tib_exact1} and \ref{fig:tib_exact2}.

\textbf{Example 1 (Figure \ref{fig:tib_exact1})} The standard-cell netlist is shown in Figure \ref{fig:tib_exact1}. Signals $data_{0}$ and $data_{1}$ are the two inputs to a 2-to-1 multiplexer. Signals \textit{a, b, c, d}, and \textit{e} are the primary inputs. In each logic cone, the first two levels logic includes one AOI21 and three NAND2 gates. Each gate is considered as a vertex. The determination process starts with $g_{0}$ and $g_{5}$. Then, two vectors of vertices are created using breath-first-search since $g_{0}$ and $g_{5}$ are the same type vertices. $V_{0}$=\{$g_1$, $g_4$\}, $V_{1}$=\{$g_7$, $g_6$\}. To maintain the traversed graphs in the isomorphic class, there exists only one pairing choice, i.e. ($g_1$, $g_{6}$), ($g_4$, $g_7$). The two vectors will be updated, $V_{0}$=\{$g_2$, $e$, $g_3$\} and $V_{1}$=\{$g_2$, $e'$, $g_3$\}. Since $x$ and $y$ are primary inputs, they are paired and eliminated from $V_{0}$ and $V_{1}$. Hence, we have two NAND2 vertices in each vector, which has two pairing options, i.e. ($g_2$, $g_8$) or ($g_2$, $g_9$). However, in the standard-cell based DAG, only one option remains. This is because AOI21 has two types of inputs, including two inputs for AND and one input for OR/NOR. To maintain the function equivalence, $g_2$ must pair with $g_8$, and so $g_3$ must pair with $g_9$. In summary, the total number of possible attempts for determining isomorphism for the first two level logic is \underline{\textbf{one}}. 

However, this approach requires much more effort to determine the maximum isomorphism while using AIG representation. The AIG representation of this design is shown in Figure \ref{fig:tib_exact2}. According to the algorithm proposed in \cite{cunxiyu:dac16}, the first level logic has two options for pairing, i.e. node $2$ with node $9$, or node $2$ with node $10$. The algorithm solves this problem using a look-ahead heuristic, which traverses three levels deeper and picks the pairing that gives more common logic. This situation happens also while checking (node $4$ with node $7$, and node $11$ with node $13$), and (node $6$ with node $7$, and node $13$ with node $14$). This means that it requires three times look-ahead checking and total of \underline{\textbf{eight}} attempts to identify the same common logic as the one shown in Figure \ref{fig:tib_exact1}. 

\subsubsection{Including side fanout information}

Based on the observation shown in Example 1, we can see that providing various types of vertices at each logic level can significantly reduce the total number of pairing attempts for isomorphism determination. Thus, we preserve the fanout information of the standard cells in the vertices. This can significantly improve the runtime for a large design that includes many reconvergent fanouts, such as the optimized multipliers. 

\begin{figure}[!htb] 
\begin{center}
\includegraphics[scale=0.28]{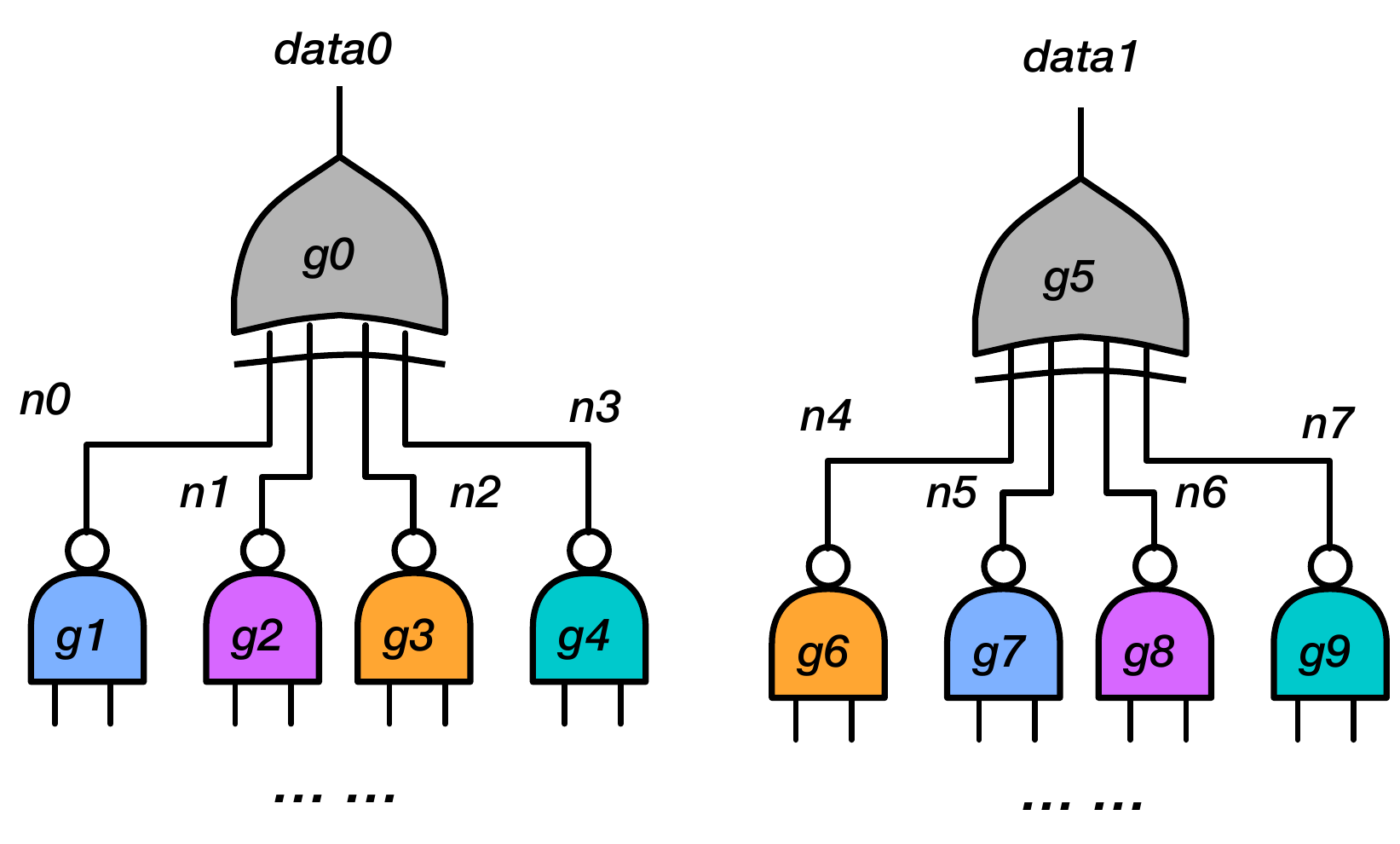}
\caption{Illustrative example of utilizing fanout information of each vertex.}
\vspace{-2mm}
\label{fig:fanout}
\end{center}
\end{figure}

\begin{figure}[!htb] 
\begin{center}
\includegraphics[scale=0.28]{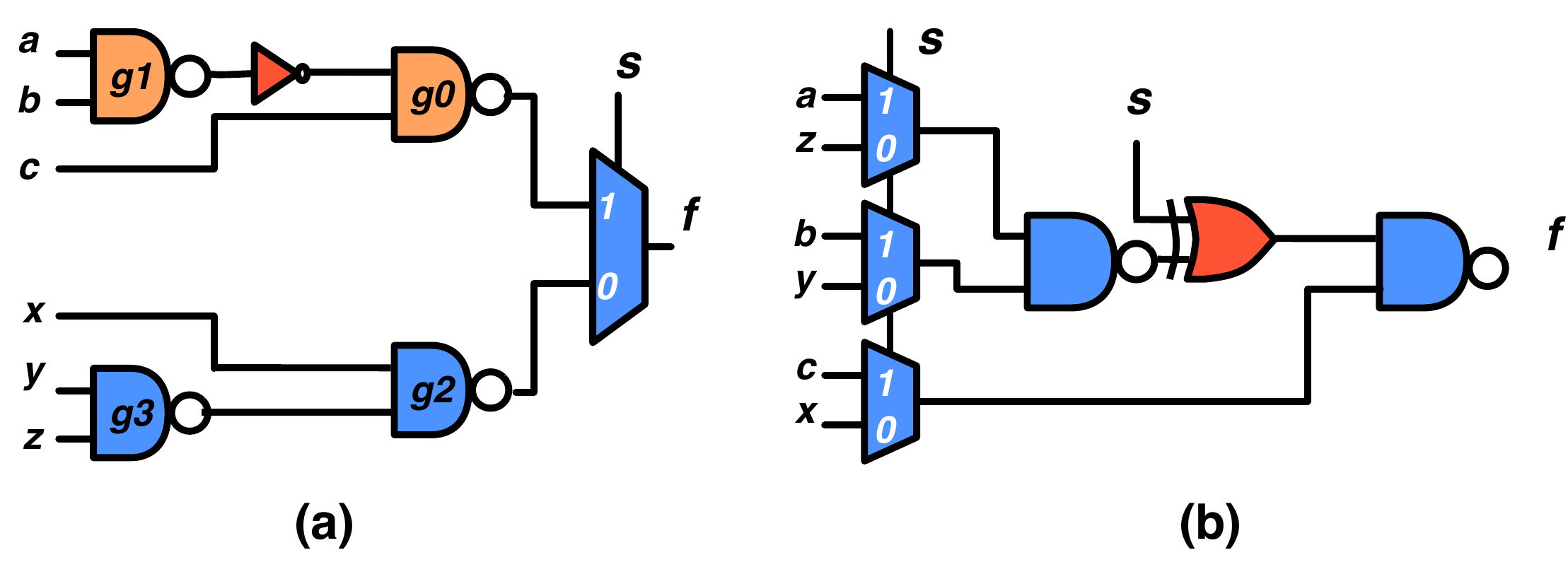}
\caption{Approximate isomorphism determination by ignoring inverters. a) original design; b) optimized design using extra XOR2 gate.}
\vspace{-2mm}
\label{fig:inv2xor}
\end{center}
\end{figure}

\textbf{Example 2 (Figure \ref{fig:fanout})} Assume that each logic cone of a 2-to-1 multiplexer includes one XOR4 and four NAND2 gates in the first two levels. Let the number of side fanouts of nets \{$n_0$, $n_1$, $n_2$, $n_3$\} be \{3,2,1,0\}, and the number of side fanouts of nets \{$n_4$, $n_5$, $n_6$, $n_7$\} be \{1,3,2,0\}. Without including the fanout information, the total number of possible pairing is $24$ since four vertices in the second level are identical. However, if we consider to pair the vertices according to the number of side fanouts, there will be only one pairing choice, i.e. ($g_1$, $g_7$), ($g_2$, $g_8$), ($g_3$, $g_6$), and ($g_4$, $g_9$). Although, the fanout information can significantly reduce the number of pairing, such case may not always exist. If so, our approach will go through the look-ahead heuristic pairing process.

  \begin{figure*}	
	\centering
	\begin{subfigure}[t]{0.3\textwidth}
		\centering
		\includegraphics[width=\textwidth]{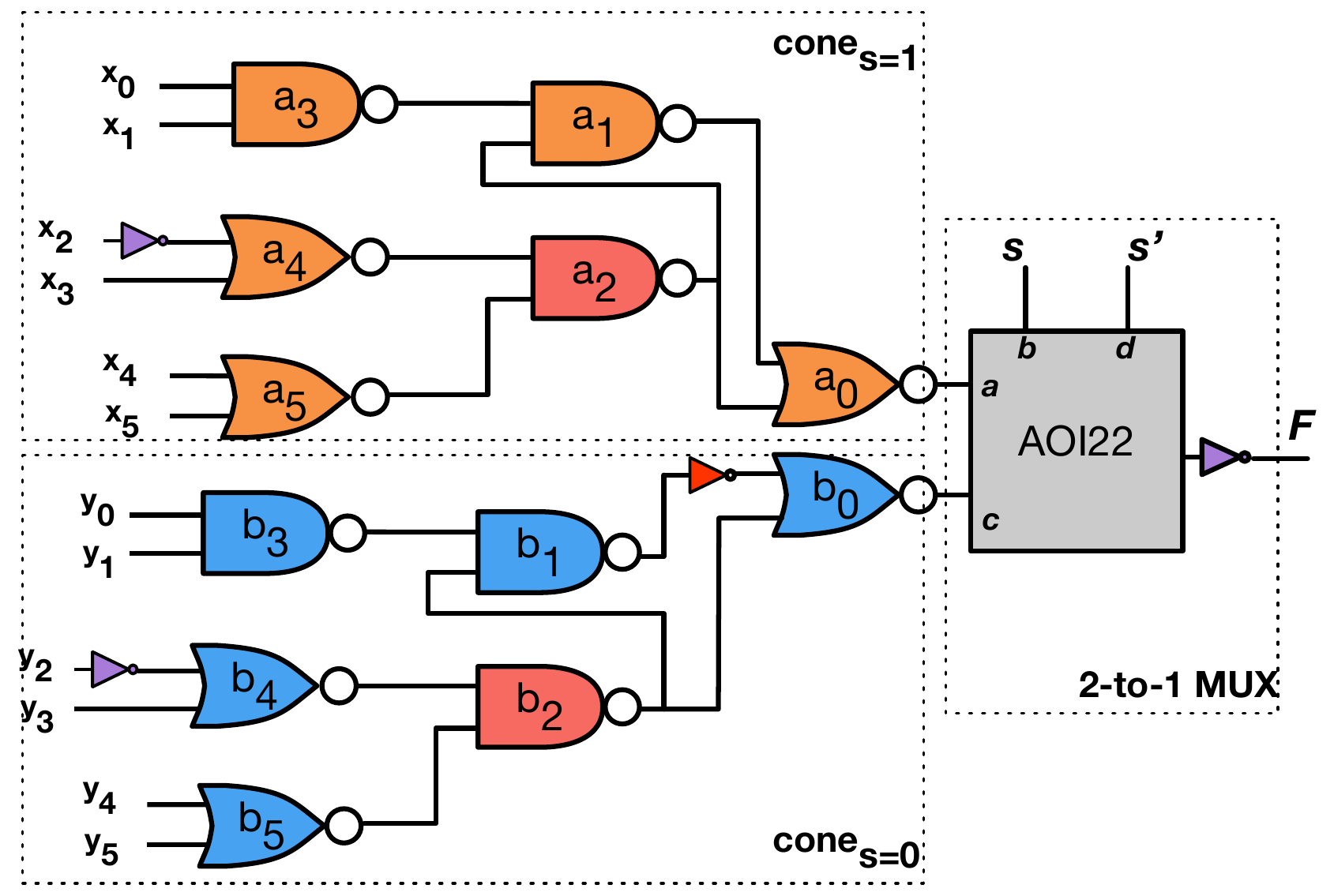}
		\caption{Original circuit.}		
	\end{subfigure}
	\quad
	\begin{subfigure}[t]{0.3\textwidth}
		\centering
		\includegraphics[width=\textwidth]{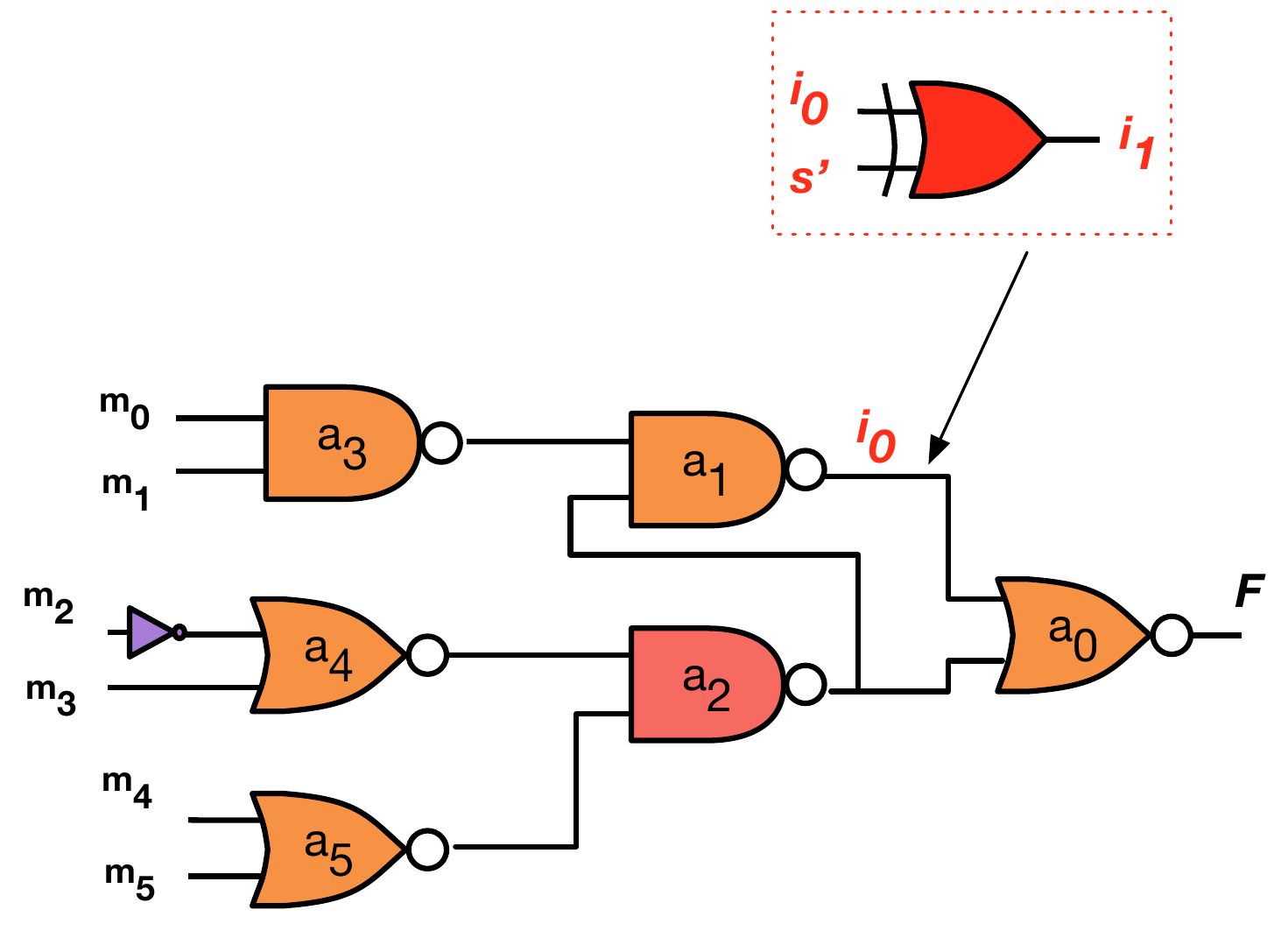}
		\caption{Circuit optimized with our approach.}
	\end{subfigure}
	\caption{A complete example of multiplexer relocation using the proposed approach.}
	\vspace{-3mm}
	\label{fig:overview-example}
	\end{figure*}

\subsection{Approximate Isomorphism Determination}

In addition to considering the exact isomorphism graph as common specification logic, a novel approximate isomorphism determination approach is developed in this work. One observation is that much more common logic exists by ignoring the inversions. For example, in the case of a 2-to-1 multiplexer that selects \textit{less than} operator and \textit{less than or equal to}, there is no common logic that can be identified using both representations while considering inversions. Thus, we propose an approximate isomorphism method to overcome this limitation. Specifically, in the process of identifying common logic, the inverters will be replaced by a 2-input XOR, with an extra input coming from the control signal of the multiplexer, or its complement.

\textbf{Example 3 (Figure \ref{fig:inv2xor})} The original netlist is shown in Figure \ref{fig:inv2xor}(a). Using the approach described in Section 3.1, there will be only one gate in each instance of the common logic, namely $g_0$ and $g_2$. However, we can see that the two logic cones connected to the 2-to-1 multiplexer are identical without considering the inverter. Hence, we continue searching for the common logic by skipping the inverters. In this example, the common logic includes two NAND2 and one inverter. To maintain the original function of $f$, the inverter is replaced by an XOR2, whose extra input is the control signal $s$. In Figure \ref{fig:inv2xor}(b), signal $s$ in the XOR2 actually selects the XOR2 to be a inverter or wire, i.e. when $s=1$, XOR2 is a inverter; and when $s=0$, XOR2 is a buffer.

\subsection{Implementation}

The implementation of single multiplexer relocation is shown in Algorithm 2. The multiplexer relocation function of sub-circuit with a vector multiplexer at the primary outputs (line 5 in Algorithm 1), is applying the single relocation function iteratively on each output bit. The input of Algorithm 2 is a sub-circuit with single output bit that is generated by a of 2-to-1 multiplexer. Algorithm 2 operated in three steps:

\begin{algorithm}
\scriptsize
\caption{Single Multiplexer Relocation}\label{alg:algorithm}
\textbf{Input: Pre-processed sub-circuit $C$}\\ 
\textbf{Output: An optimized standard-cell netlist}\vspace{2mm}\\
\textbf{Single\_Mux\_Relocate($C$)}
\begin{algorithmic}[1]
\State $B$ = RelocationBoundray($PO$)
\State $C$ $\gets$ relocate multiplexer to level $B$, w/o considering inverters
\State $P$ = inv2xorPosition($PO$, $B$)
\State $C$ $\gets$ insert XORs to $P$ based on its location \\
\Return $C$
\end{algorithmic}
\textbf{RelocationBoundray($PO$)}
\begin{algorithmic}[1]
\State $m \gets levels(PO) - 1$; \/\/inverter is considered as $0$ level.
\While{$m \ge 0$}
\State $L0_{m} \gets$ the gates in ($s=0$) logic at level $m$
\State $L1_{m} \gets$ the gates in ($s=1$) logic at level $m$
%\If{($L0'_{m}, L1'_{m}$) $\gets$ uniqueFanoutPairs($L0_{m}, L1_{m}$)}
\If{uniqueFanoutPairs($L0_{m}, L1_{m}$)}
	\State $U0_{m}, U1_{m} \gets$ uniqueFanoutPairs($L0_{m}, L1_{m}$)
	\State $L0_{m} \gets$ $L0_{m}$ - $U0_{m}$; $L1_{m} \gets$ $L1_{m}$ - $U1_{m}$
	\State $L0_{m+1}, L1_{m+1}$ $\gets$ ($U0_{m}, U1_{m}$)+isomorphism($L0_{m}, L1_{m}$)
\Else
	\If {isomorphsim($L0_{m}, L1_{m}$)}
		\State $L0_{m+1}, L1_{m+1} \gets$ isomorphsim($L0_{m}, L1_{m}$)
	\Else
		\State{ \textbf{exit}} %(level(PO) - $1$ - $m$), ($L0_{m-1}, L1_{m-1}$)}
	\EndIf
%\Else
%\Return level(PO) - $1$ - $m$
\EndIf
\EndWhile \\
\Return (level(PO) - $1$ - $m$), ($L0_{m-1}, L1_{m-1}$)
\end{algorithmic}

\textbf{inv2xorPosition($PO$, boundary)}
\begin{algorithmic}[1]
\State $P_{0} \gets$ the positions of all inverts up to $boundary$ level
\State $P_{1} \gets$ the positions of all inverts up to $boundary$ level \\
\Return $P_{0} \cap P_{1}$ 
\end{algorithmic}
\end{algorithm}

\noindent
\textbf{a)} The key function of this approach is identifying the maximum common specification logic connected to the multiplexer. The function is described in function \textbf{RelocationBoundray} in Algorithm 2. Specifically, our algorithm identifies the boundary logic cut where the isomorphism between two logic cones ends. This function also returns the pairings of the boundary signals that maintains the isomorphism class, which is used for creating the new multiplexers.  

We backward traverse the graph from the two inputs of the 2-to-1 multiplexer level by level (lines 1 - 2). The gates at level $m$ are stored in two vectors (lines 3 - 4), depending their selecting signal. As mentioned in Section 3.1.2, our approach benefits signicantly from the fanout information. Hence, we first check if there exist unique fanout pairs. If so, we eliminate those pairs from the two vectors that store the gates. The rest of the gates in the two vectors will do a regular isomorphism check, with a 3-depth look ahead search \cite{cunxiyu:dac16}. For example, in Figure \ref{fig:overview-example}, there are two NAND2 gates in each vector, ($a_{1}$, $a_{2}$) and ($b_{1}$, $b_{2}$). There are two pairing choices at this level, i.e., ($a_{1}$, $b_{1}$) and ($a_{2}$, $b_{2}$), or ($a_{1}$, $b_{2}$) and ($a_{2}$, $b_{1}$). Using the fanout information, there will be only one feasible pairing, i.e., ($a_{1}$, $b_{1}$) and ($a_{2}$, $b_{2}$). This is because $a_{2}$ and $b_{2}$ have two fanouts, and $a_{1}$ and $b_{1}$ have only one fanout.

\noindent
\textbf{b)} Relocate the multiplexer across the common specification logic, up to the boundary cut returned by the previous step. The two logic cones between boundary and the multiplexer output have common specification (not functionally equivalent), denoted as $cone_{s=0}$ and $cone_{s=1}$, depending on the select signal of the multiplexer. To relocate the multiplexer, we disconnect all the pins of $cone_{s=1}$ and create a set of multiplexers that select the inputs signals of those two logic cones. For example, in Figure \ref{fig:overview-example}, $m_{i}$=$x_{i}{s}$+$y_{i}\bar{s}$, $i$=\{1,2,3,4,5\}. Then, the inputs of $cone_{s=1}$ will be replaced by the outputs of the new multiplexers. In this case, $x_{i}$ is replaced by $m_{i}$. Finally, the output $F$ will be reconnected to the output of $cone_{s=1}$. 

\noindent
\textbf{c)} In the function of \textbf{RelocationBoundray}, we do not consider inverter as a gate, or a node in the DAG. This enables the approximate isomorphism determination (Section 3.2). As mentioned earlier, this allows us to identify a larger common logic. For example, if we consider inverter as a node in the graph, the common logic will consist of only two NOR2 gates, $a_{0}$ in $cone_{s=1}$ and $b_{0}$ in $cone_{s=0}$. To maintain the functionality of the design, we need to insert XOR2 gates with extra input $s$ or $\bar{s}$ depending on which cone the invert belongs to. We first record the locations of all inverters in $cone_{s=0}$ and $cone_{s=1}$, denoted as $P_{0}$ and $P_{1}$, up to the boundary cut. The locations that require an XOR2 replacement is included in the result of $P_{0} \cap P_{1}$. This is why the inverters connected to gates $a_{4}$ and $b_{4}$ do not require XOR2 insertion, since they maintain the two cones in the isomorphism class (Figure \ref{fig:overview-example}). The inverter connected to $b_{0}$ requires an XOR2 insertion, and it belongs to $cone_{s=0}$. Hence, an XOR2 with extra input $\bar{s}$ is inserted to replace $i_{0}$ in Figure \ref{fig:overview-example}.

%\input{demo}
% Please add the following required packages to your document preamble:
\begin{table*}[]
\scriptsize
\centering
\begin{tabular}{|c|r|r|r|r|r|r|}
\hline
\multirow{2}{*}{(n-bit) Operators} & \multicolumn{2}{c|}{IBM flow} & \multicolumn{2}{c|}{\begin{tabular}[c]{@{}c@{}}IBM flow\\ with AIG Opt\end{tabular}} & \multicolumn{2}{c|}{\begin{tabular}[c]{@{}c@{}}IBM flow\\ with \textbf{our approach}\end{tabular}} \\ \cline{2-7} 
 & Area & Lev & Area & Lev & Area & Lev \\ \hline
(64), A$<$B:A$<$C & 2280 & 11 & 2124 & 13 & 1855 & 15 \\ \hline
(64), A+B, A+C & 10162 & 17 & 9333 & 15 & 5787 & 20 \\ \hline
(64), A+B:A-C & 8697 & 19 & 8104 & 25 & 8062 & 21 \\ \hline
(64) A$<$B:A$<=$B & 2464 & 12 & 2126 & 13 & 2198 & 12 \\ \hline
*(64) A$\times$B:A$\times$C & 182917 & 83 & 482811 & 211 & 91245 & 89 \\ \hline
A$\times$B$/$C{[}7:0{]}:A$\times$B$/$C{[}15:8{]} & 3626 & 26 & 5606 & 26 & 1760 & 27 \\ \hline
(32) A$\times$B+C:B$\times$C+A & 52943 & 58 & 108402 & 120 & 26709 & 58 \\ \hline
(6) dec(A):dec(B) & 1319 & 5 & 667 & 5 & 549 & 7 \\ \hline
 & 1 & $+$0 lev & 1.106 & $+$1.16 lev & \textbf{0.658} & $+$2 lev \\ \hline
\end{tabular}
\caption{Results of arithmetic test cases using the original IBM synthesis Flow, IBM synthesis flow with AIG optimization, and original IBM synthesis flow with the proposed approach.(*This design is not used for comparison.)}
\vspace{-2mm}
\label{my-label}
\end{table*}
\begin{table*}[]
\scriptsize
\centering
\begin{tabular}{|c|r|r|r|r| |r|r|r|r|}
\hline
\multirow{2}{*}{Benchmarks} & \multicolumn{2}{c|}{\begin{tabular}[c]{@{}c@{}}Flow1\end{tabular}} & \multicolumn{2}{c|}{\begin{tabular}[c]{@{}c@{}}Flow1 with\\our approach\end{tabular}} & \multicolumn{2}{c|}{\begin{tabular}[c]{@{}c@{}}Flow2\end{tabular}} & \multicolumn{2}{c|}{\begin{tabular}[c]{@{}c@{}}Flow2 with\\our approach\end{tabular}} \\ \cline{2-9} 
 & Area & Delay & Area & Delay & Area & Delay & Area & Delay \\ \hline
ibm1 & 3622 & 216.45 & 2223 & 255.81 & 4587 & 295.16 & 2235 & 255.81 \\ \hline
ibm2 & 5454 & 314.84 & 3361 & 354.19 & 6879 & 432.90 & 3366 & 383.71 \\ \hline
ibm3 & 9115 & 501.77 & 5526 & 501.77 & 11463 & 688.71 & 5610 & 541.13 \\ \hline
ibm4 & 12782 & 678.87 & 7874 & 649.35 & 16047 & 924.84 & 7854 & 688.71 \\ \hline
ibm5 & 18323 & 924.84 & 11121 & 787.10 & 22923 & 1200.32 & 12342 & 875.65 \\ \hline
ibm6 & 27435 & 1170.81 & 16843 & 983.87 & 34383 & 1505.32 & 16803 & 1023.23 \\ \hline
ibm7 & 31069 & 1288.87 & 19083 & 1082.26 & 38967 & 1603.71 & 19074 & 1082.26 \\ \hline
 & 1 & 1 & \textbf{0.613} & \textbf{0.970} & 1 &1  & \textbf{0.487} & \textbf{0.767} \\ \hline
\end{tabular}
\caption{Evaluation of our approach in the complete production Flow using industry designs in 14nm technology. Flow1 is the IBM synthesis flow with AIG optimization; Flow2 is the original IBM synthesis flow.}
\vspace{-3mm}
\label{my-label}
\end{table*}

\section{Experimental results}

The proposed approach in this Section 3 was implemented in C++ and integrated with the IBM logic synthesis flow \cite{stok1996booledozer} and further evaluated with IBM high-level synthesis flow and Place and Route (P\&R) flow. Our approach is performed before technology mapping within the logic synthesis flow. The program was tested on a number of datapath designs in SystemC. The datapath designs include large arithmetic operators, such as 64-bit multipliers. All the experimental results are collected at the end of the complete production design flow. This demonstrates that our approach successfully overcomes the limitations of the existing logic synthesis and high-level synthesis techniques reviewed in Section 1. All of our experimental results are obtained using high-performance 14nm technology library. To demonstrate the runtime improvement compared to the work of \cite{cunxiyu:dac16}, we examine the runtime using a set of designs, including a multiplier circuit up to 64 bits. Our experiments were conducted on a machine with Intel(R) Xeon CPU 7560 v6 2.20 GHz x32 with 4 TB memory.

\begin{figure}[!htb] 
\begin{center}
 \includegraphics[width=0.35\textwidth]{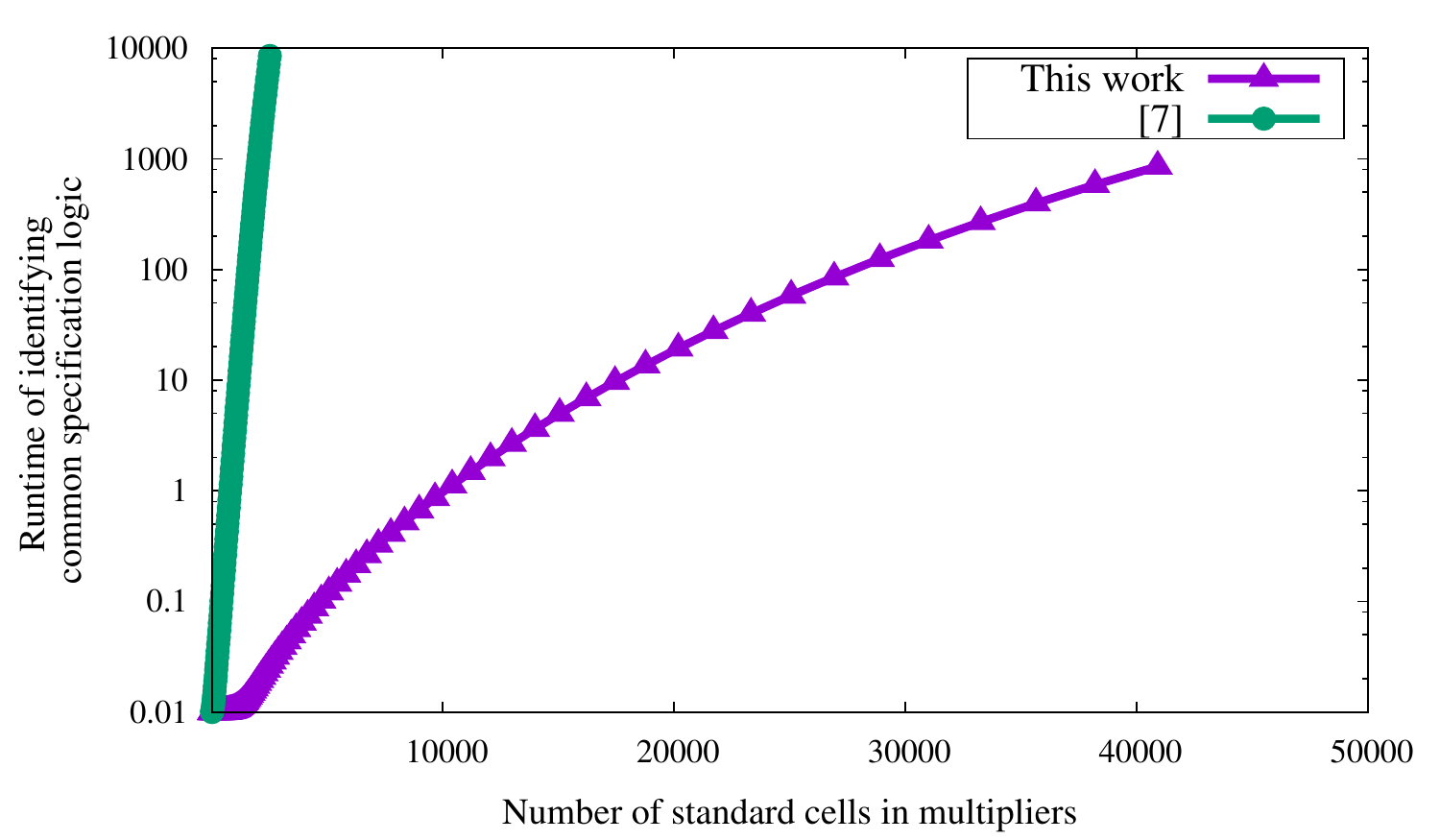}
\caption{Evaluation of CPU runtime using designs with multipliers compared to \cite{cunxiyu:dac16}.}
\vspace{-3mm}
\label{fig:comp}
\end{center}
\end{figure}

We first evaluate our approach using a set of arithmetic designs in which there are two arithmetic operators selected by control signals. The results are shown in Table 1. The first column indicates the bit-width of the arithmetic operators and the type of the two operators. These designs are implemented in SystemC using \textit{''if then else''} statement. The second and third columns show the area and logic level results produced by the original IBM synthesis flow. The fourth and fifth columns show the results produced by the original flow with combinational AIG optimization \cite{mishchenko2010abc}. The last two columns show the results produced by original flow with our approach. The last row shows the average improvement gain or loss. Specifically, the increase or decrease area is measured in percentage of the original flow, and the change of logic level is measured in the number of levels. Based on Table 1, we can see that: 1) our approach gives on average 34\% area reduction compared to the other two flows. Note that the flows include complete high-level and logic-level optimizations techniques; and 2) our approach can handle large complex arithmetic operators, such as datapath with large multipliers. With approximate isomorphism determination, we can optimize the design with various combinations of two different operators.

We then evaluate our approach using seven industrial designs implemented in SystemC. Two synthesis flows are used for experiments: \textit{Flow1} is the IBM synthesis flow with AIG optimization; \textit{Flow2} is the original IBM synthesis flow. The results are shown in Table 2. The second and third columns show the results produced by Flow1, and fourth and fifth columns are produced by Flow1 with our approach. The sixth to seventh columns show the results produced by Flow2. We compare the average improvement of the area and the delay at the last row. We can see that both area and delay have been improved in these experiments. Specifically, using Flow1 the area on average reduces by 39\%, and the delay on average reduces 3\%, and Flow2 offers 51\% area reduction with 23\% delay improvement on average. Note that the delay improvements are not provided directly by our approach. The delays are improved because our approach enables other optimization techniques. Specifically, for those benchmarks, an Adder optimization technique \cite{roy2014towards} implemented in the IBM synthesis flow is enabled and significantly improves the delay after relocating the multiplexers.

\begin{table}[!htb]
\scriptsize
\centering
\caption{Comparing the PnR results with multiplexer relocation with the original flow.}
\label{tbl:PnR}
\begin{tabular}{|c|c|c|c|}
\hline
Benchmarks & \multicolumn{1}{c|}{Route Length} & \multicolumn{1}{l|}{Power} & \multicolumn{1}{c|}{Worst-case delay} \\ \hline
ibm1 & \textbf{0.73} & \textbf{0.45} & \textbf{0.95} \\ \hline
ibm2 & \textbf{0.79} & \textbf{0.61} & \textbf{0.97} \\ \hline
ibm4 & \textbf{0.92} & \textbf{0.71} & 1.06 \\ \hline
ibm6 & 1.23 & \textbf{0.78} & 1.10 \\ \hline
\end{tabular}
\vspace{-3mm}
\end{table}

  \begin{figure*}	
	\centering
	\begin{subfigure}[t]{0.3\textwidth}
		\centering
		\includegraphics[width=\textwidth]{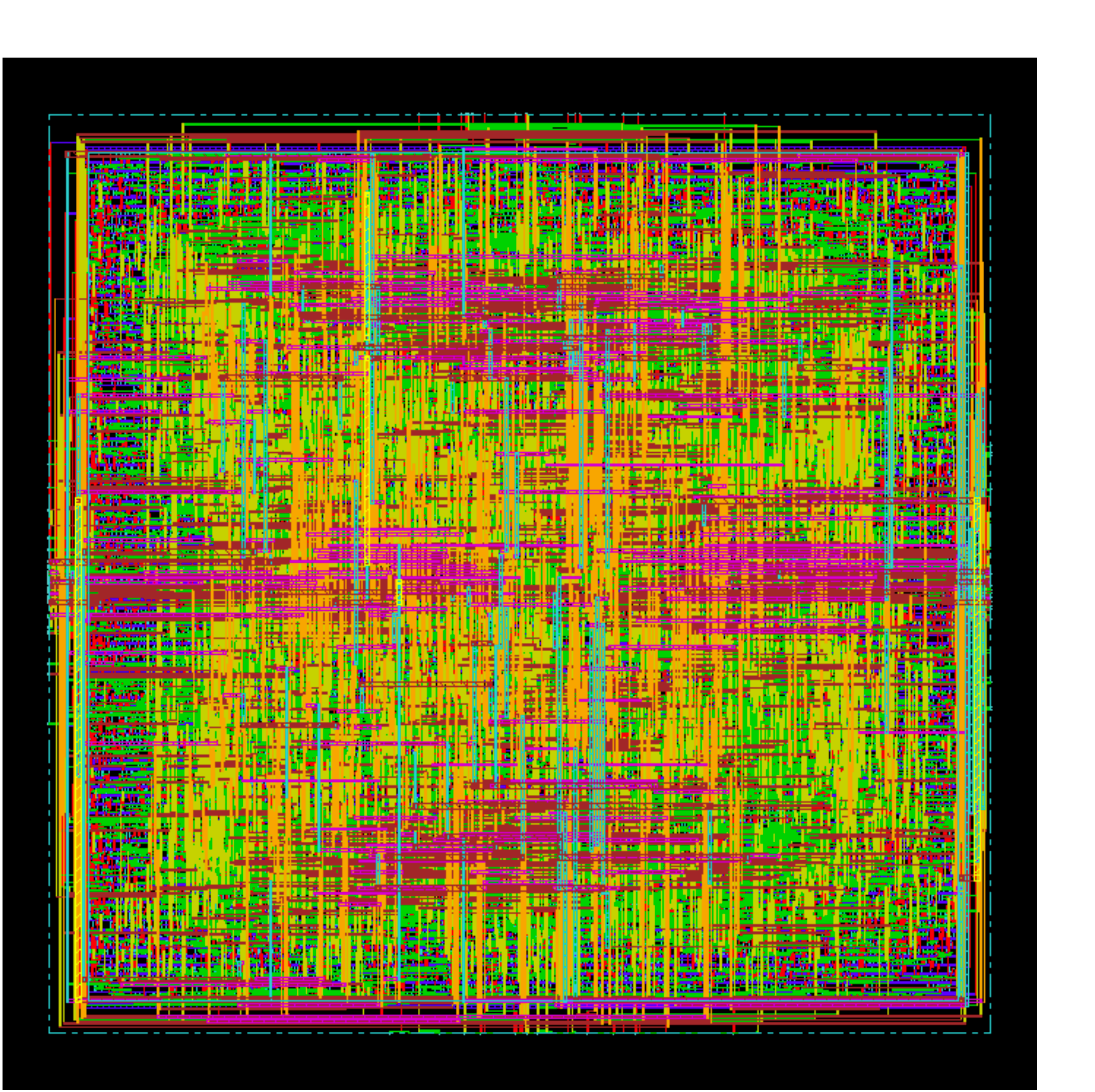}
		\caption{P\&R result of design \textit{ibm2} without multiplexer relocation.}		
	\end{subfigure}
	\quad
	\begin{subfigure}[t]{0.3\textwidth}
		\centering
		\includegraphics[width=\textwidth]{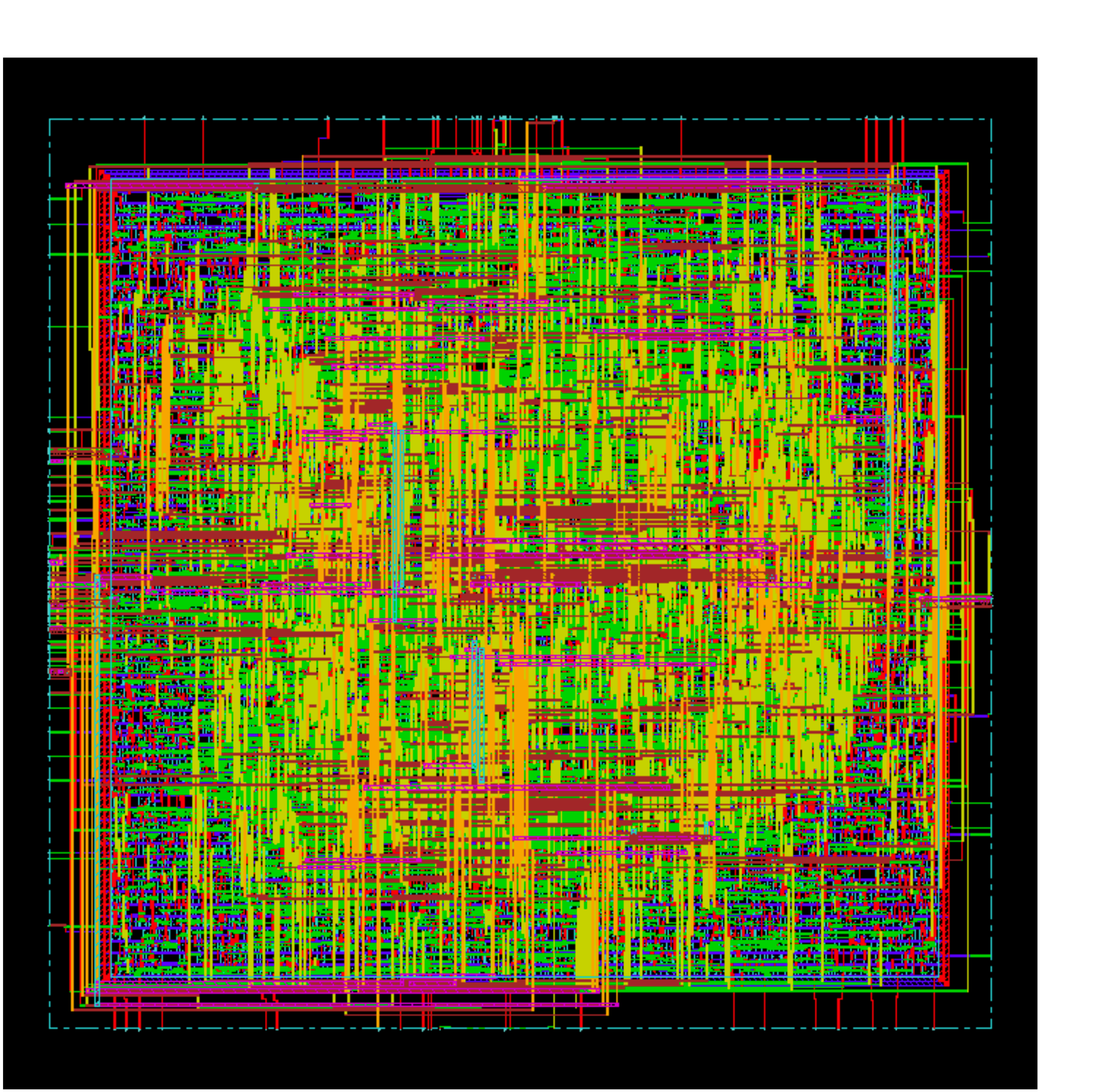}
		\caption{P\&R result of design \textit{ibm2} with multiplexer relocation.}
	\end{subfigure}
	\caption{Comparing the P\&R results using design \textit{ibm2} with and without our approach.}
	\vspace{-3mm}
	\label{fig:PnR-ibm2}
	\end{figure*}

Additionally, we evaluate our approach using four designs, \textit{ibm1, ibm2, ibm4, ibm6}, with placement and route (P\&R). The inputs of P\&R process are the designs produced by Flow1 with AIG optimization ($4^{th}$ and $5^{th}$ columns in Table 2). The routing length, power and worst-case delay are included in Table \ref{tbl:PnR}. The improvements of the area of placing the standard cells remain the same as shown in Table 2 with the same density. The P\&R results of \textit{ibm2} are shown in Figure \ref{fig:PnR-ibm2}. We can see that except \textit{ibm6}, the designs are improved successfully using our approach without delay overhead. Particularly, we observe that the power has been significantly improved compared to the original designs. Moreover, we can see that the improvements of \textit{ibm4} and \textit{ibm6} gained after P\&R are less than in the other two designs. The possible reasons for that are: 1) there are large ($\geq$32) fanout signals generated by multiplexer relocation in those two designs; and 2) a large number of the extra multiplexers have been placed tightly, which decreases routability.

The reason why we didn't compare our approach to the work of \cite{cunxiyu:dac16} in the experiments shown in Table 1 and Table 2 is the following: 1) that algorithm can't be successfully applied on all of the design within eight hours; and 2) for the designs that on which the algorithm runs successfully, the results are worse, e.g., 3rd and 4th designs in Table 1. To demonstrate that our approach significantly improves the CPU runtime compared to the existing algorithm in the cases of datapaths with \underline{multipliers}, the experimental results are provided in Figure \ref{fig:comp}. %, using two sets of designs where: 1) two additions are selected; 2) two multiplications are selected. Both algorithms can successfully identify the maximum common specification logic, and relocate the multiplexers for the first set of designs up to 256-bit within a minute. 
%The runtime of the multiplication-based design is very different. 
The designs used for the experimental results shown in Figure \ref{fig:comp} vary from 4-bit to 64-bit. In Figure \ref{fig:comp}, the x-axis represents the number of standard cells in the design, and the y-axis represents the CPU runtime of the multiplexer relocation algorithm in logarithmic scale. It is clear that our algorithm performs much faster than the AIG-based algorithm \cite{cunxiyu:dac16}.

\section{Conclusion}

This paper presents an advanced DAG-based algorithm that targets area minimization using logic-level resource sharing. The common specification logic identification is formulated as \textit{unweighted} graph isomorphism problem. In addition, an \textit{approximate} isomorphism algorithm is proposed in this paper to identify extra common logic. The proposed approach demonstrates that it can significantly reduce area, and potentially reduce delay on industrial designs, within a complete design flow. The runtime has been reduced from exponential to linear comparing to the existing algorithms. Future work will focus on improving function of identifying common specification logic.

\vspace{-2mm}

\bibliographystyle{IEEEtran}
\bibliography{verification_ycunxi.bib}
%\bibliography{/Users/cunxiyu/Documents/TexShop/cunxi_bib/verification_ycunxi}
\end{document}